\documentclass[fleqn,usenatbib]{mnras}
\usepackage{newtxtext,newtxmath}

\usepackage[T1]{fontenc}

\DeclareRobustCommand{\VAN}[3]{#2}
\let\VANthebibliography\thebibliography
\def\thebibliography{\DeclareRobustCommand{\VAN}[3]{##3}\VANthebibliography}

\usepackage{graphicx}	

\newcommand\un[1]{{\,\rm #1}}
\newcommand\E[1]{\times10^{#1}}
\newcommand\rs[1]{_\mathrm{#1}}

\usepackage[dvipsnames]{xcolor}

\title[Magnetic field in SN1987A]{Polarized radio emission unveils the structure of the pre-supernova circumstellar magnetic field and the radio emission in SN1987A}

\author[O. Petruk et al.]{%
O. Petruk,$^{1,2,3}$\thanks{E-mail: oleh.petruk@gmail.com}
V. Beshley,$^{1}$
S. Orlando,$^{3}$
F. Bocchino,$^{3}$ 
M. Miceli,$^{4,3}$
\newauthor
S. Nagataki,$^{5,6}$
M. Ono,$^{5,6}$
S. Loru,$^{7}$
A. Pellizzoni,$^{8}$
E. Egron$^{8}$
\\
$^{1}$Institute for Applied Problems in Mechanics and Mathematics, Naukova St. 3-b, 79060 Lviv, Ukraine\\
$^{2}$Astronomical observatory, Ivan Franko National University of Lviv, Kyryla i Methodia St. 8, UA-79005 Lviv, Ukraine\\
$^{3}$INAF - Osservatorio Astronomico, Piazza del Parlamento 1, 90134 Palermo, Italy\\
$^{4}$Dip. di Fisica e Chimica, Università degli Studi di Palermo, Piazza del Parlamento 1, 90134 Palermo, Italy\\
$^{5}$Astrophysical Big Bang Laboratory, RIKEN Cluster for Pioneering Research, 2-1 Hirosawa, Wako, Saitama 351-0198, Japan\\
$^{6}$ RIKEN Interdisciplinary Theoretical \& Mathematical Science Program (iTHEMS), 2-1 Hirosawa, Wako, Saitama 351-0198, Japan\\
$^{7}$INAF - Osservatorio Astrofisico di Catania, Via Santa Sofia 78, 95123 Catania, Italy\\
$^{8}$INAF - Osservatorio Astronomico di Cagliari, Via della Scienza 5, 09047 Selargius, Italy
}

\date{Accepted XXX. Received YYY; in original form ZZZ}

\pubyear{2022}

\begin{document}
\label{firstpage}
\pagerange{\pageref{firstpage}--\pageref{lastpage}}
\maketitle

\begin{abstract}
The detected polarized radio emission from remnant of SN1987A opens the possibility to unveil the structure of the pre-supernova magnetic field in the circumstellar medium.  {Properties derived from direct measurements would be of importance for understanding the progenitor stars and their magnetic fields.} As the first step to this goal, we adopted the hydrodynamic data from an elaborated three-dimensional (3-D) numerical model of SN1987A. We have developed an approximate method for `reconstruction' of 3-D magnetic field structure inside supernova remnant on the `hydrodynamic background'. This method uses the distribution of the magnetic field around the progenitor as the initial condition. With such a 3-D magneto-hydrodynamic model, we have synthesized the polarization maps for a number of SN1987A models and compared them to the observations. In this way, we have tested different initial configurations of the magnetic field as well as a structure of the synchrotron emission in SN987A. We have recovered the observed polarization pattern and we have found that the radial component of the ambient pre-supernova magnetic field should be dominant on the length-scale of the present-day radius of SN1987A.  {The physical reasons for such a field are discussed.}
\end{abstract}



\section{Introduction}

Remnant of supernova 1987A is an attractive object 
for studies \citep[e.g.][]{2016ARA&A..54...19M}, for instance: to investigate how the structure and morphology of supernova remnants (SNRs) reflect the properties of the parent supernova (SN) explosions and the nature of the progenitor stars, to unveil how dust forms after SN explosions, to reconstruct the pre-SN circumstellar medium (CSM) and stellar magnetic field formed in the latest phases of progenitor star evolution, to understand the spatial distribution of emitting relativistic particles inside SNR. In fact, SN 1987A is the only SN exploded so close to us after the invention of the telescope. 
It is accurately monitored since the explosion event. Observations span from  radio \citep[e.g.][]{2010ApJ...710.1515Z, 2013ApJ...767...98Z, 2016MNRAS.462..290C, 2017ApJ...842L..24A, 2018ApJ...867...65C} through the optic and infra-red bands \citep[e.g.][]{2010Sci...329.1624F,2015ApJ...806L..19F,2016ApJ...833..147L,2020ApJ...890....2A, 2022MNRAS.511.2977K}
to X-rays \citep[e.g.][]{2011ApJ...733L..35P, 2013ApJ...764...11H, 2016ApJ...829...40F,2019NatAs...3..236M, 
2021ApJ...916...41S, 2021ApJ...922..140R, 2021ApJ...908L..45G, 2022ApJ...931..132G} and contain a wealth of information about the stellar ejecta, remnant and physics inside. 

High-performance three-dimensional (3-D) simulations aim to restore the explosion  \citep{2019ApJ...882...22A,2020ApJ...888..111O}, as well as the structure and expansion of SN1987A including spectral variations and detailed morphology of the remnant and ejecta, in particular in X-rays \citep{2015ApJ...810..168O,2020A&A...636A..22O} and in the radio band \citep{2014ApJ...794..174P,2019A&A...622A..73O}. The later paper reports the first 3-D magneto-hydrodynamic (MHD) numerical simulations of the remnant of SN1987A. 

Important new milestone in observations of SN1987A is detection of the polarized radio emission \citep{2018ApJ...861L...9Z}. For the first time, it allows one to test models of magnetic field (MF) in this supernova remnant (SNR). One of the most distinctive feature apparent from this observational result is the dominant radial orientation of the MF vectors. This is in line with what is found in other young SNRs \citep{1976AuJPh..29..435D,2015A&ARv..23....3D}.

The MF structure inside SNR and thus the polarization pattern depend on MF configuration in and around a progenitor in the pre-explosion era,  {on its behavior during the development of the explosion and on properties of its evolution from the shock breakout to the evolved SNR}. 
Synthesis of polarization images essentially require knowledge of the internal 3-D morphology of MF in SNR. Thus, the full 3-D MHD simulations are required in order to produce synthetic polarization maps.  {The ideal simulations reproduce subsequently as SN event as development of the remnant in three dimensions. This is a quite demanding task because the evolution covers many orders in time (from microseconds to thousand years) and space (from astronomical unit to few parsecs). Therefore, a viable approach could consist in the two phases: i) simulations of an SN event, ii) simulations of a SN remnant with the outcome from the previous step as initial conditions.

Though the MHD studies as SNe as SNRs were initiated few decades ago \citep[e.g.][]{1970ApJ...161..541L,1975MNRAS.172...55F}, the explorations of 3D MHD models is quite new field.  
There are two directions in MHD studies of SN explosions \citep[see an overview by][]{2022arXiv220611914R}. Namely, the magneto-rotational core-collapse SNe \citep{2015Natur.528..376M,2018JPhG...45h4001O,2020ApJ...896..102K,2020MNRAS.492.4613O,2021MNRAS.503.4942O,2021MNRAS.500.4365A} and neutrino-driven core-collapse SN explosions of non-rotating stars \citep{2020MNRAS.498L.109M,2022arXiv220411009V,2022MNRAS.516.1752M}.
It is shown in particular that MF provides a crucial support for development of explosions in both these types of models and the   
probability to explode lowers for models with low MF.

We have studied the role and behavior of MF during different phases of SNR evolution \citep{2016MNRAS.456.2343P,2018MNRAS.479.4253P,2021MNRAS.505..755P} and have shown that MF is dynamically important in SNRs only in the post-adiabatic epochs. It is obvious however that the synchrotron images and polarization patterns depend on the structure of MF inside SNR at any evolutionary stage \cite[e.g.][]{2007A&A...470..927O,2017MNRAS.470.1156P}. 

In our paper, we would like to get hints about the pre-explosion MF from observations of polarization of SN1987A. Such constraints cast light on the relatively uncertain models of the evolution of massive stars, their explosions and supernova remnants. 
Determination of the pre-SN MF configuration of the progenitor star is important to obtain information, in general, on the latest phases of evolution of massive stars and, for this particular case, to obtain information on the evolution of a binary system evolving through a common envelope phase as suggested by some authors \citep[e.g.][]{2007Sci...315.1103M}.

Up to now, there is no MHD model for the explosion of SN1987A. Therefore, an \textsl{ad hoc} configuration of magnetic field was introduced in the MHD model for the remnant of this supernova     \citep{2019A&A...622A..73O}.
In Sect.~\ref{sn87aMF:sect4} of the present paper, we analyse the polarization patterns as they appear in this model.} 

In order to check different ideas about initial MF configuration and explore the parameter space, one needs to run massive 3-D MHD simulations for each eventual MF model and parameter set. 
Such simulations are quite demanding because they should be performed in highest possible resolution. In the present paper, we adopt a different approach.
The existing hydrodynamic (HD) model of SN1987A \citep{2015ApJ...810..168O,2019A&A...622A..73O,2020A&A...636A..22O} agrees well with a wealth of observational information. 
Therefore, a method for reconstruction of MF structure inside SNR on top of the known 3-D HD background would be useful because it allows one to test a given MF model by comparison of synthetic polarized images with the observations without massive 3D simulations. 
The details of such an approximate approach are described in Sect.~\ref{sn87aMF:sect2} where we also test the method and show its limitations. The method is applied to SN1987A and results are discussed in Sect.~\ref{sn87aMF:sect3}.

\section{Polarization maps of SN1987A from the numerical MHD model}
\label{sn87aMF:sect4}

\subsection{Observed polarization and our approach}

The radio emission from SN1987A is monitored since the very beginning of the remnant evolution. The polarized intensity is a minor fraction of the total intensity. The later grows with time \citep{2010ApJ...710.1515Z} and the polarized  {intensity -- being a fraction of the total -- is expected to grow as well}. In fact, the polarized radio emission from SN1987A was detected, for the first, time with Australia Telescope Compact Array \citep{2018ApJ...861L...9Z}. 
Fig.~3 in this reference reports the image at 22 GHz which is of interest for our study. 
The rotation measure in SN987A is compatible with zero at this frequency \citep[Fig.~2 in][]{2018ApJ...861L...9Z} and the pattern of polarization vectors shows the MF directions unchanged by the Faraday effect. Error in the polarization angles is reported to be less than 5\%.

The first 3-D MHD model of SN1987A as a result of self-consistent massive numerical simulations was reported by \citet{2019A&A...622A..73O}. The model of supernova used as initial conditions for the SNR evolution was 1-D; it was remapped in the 3-D domain in a spherically symmetric way. In the following study \citep{2020A&A...636A..22O} the model was elaborated in order to include the ejecta asymmetry derived in dedicated 3-D simulations of the core-collapse SN explosion \citep{2020ApJ...888..111O}. 

In the present section, we use the model MOD-B1 \citep[described in Sect.~2.1 in][]{2019A&A...622A..73O}.  
It assumes the magnetic field configuration to be the common \citet{1958ApJ...128..664P} spiral
\begin{equation}
    B\rs{r}=\frac{A_1}{r^2},\qquad B\rs{\phi}=-\frac{A_2}{r}\sin\theta
    \label{sn87a:parker1}
\end{equation}
with
\begin{equation}
    A_1=B\rs{0}r\rs{0}^2,\qquad A_2=B\rs{0}r\rs{0}^2\omega\rs{s}/u\rs{w},
    \label{sn87a:parker2}
\end{equation}
where $B\rs{0}$ is the MF strength at the surface of the star, $r\rs{0}$ its radius, $\omega\rs{s}$ the angular velocity of the stellar rotation, $u\rs{w}$ is the speed of wind, $\theta=\pi/2$ in the equatorial plane. Numerically, the parameters  $A_1=3\E{28}\un{G\ cm^2}$, $A_2=8\E{10}\un{G\ cm}$ were used for the MOD-B1 model \citep[see visualization of the initial MF configuration on Fig.~1 in][]{2019A&A...622A..73O}.

Numerical simulations reported by \citet{2019A&A...622A..73O} were performed on a grid with $1024^3$ zones. The data cubes were downgraded to the matrices of the size $256^3$ for image analysis in the present paper. Such resolution shortens time for production of the polarization maps while keeping all its features. 
In the present paper, we use the data for year 30 of SN1987A evolution which corresponds to epoch observed by ATCA \citep{2018ApJ...861L...9Z}. 
Images of the SNR projection on the plane of the sky which we derive correspond to the frequency 22 GHz and are smoothed to match the resolution of the observations. 

\citet{2005ApJS..159...60S} found that the structures in and around SN1987A are inclined in respect to the fixed Cartesian coordinate system ($x$-axis to the West, $y$-axis to the North, $z$-axis toward the observer) on angles $41^\mathrm{o}$ with axis $x$ (i.e. the upper side of the dense ring is closer to the observer\footnote{In this reference, the plane of the equatorial ring was considered initially to be vertical, i.e. to coincide with the $xy$ plane, with the major axis of ellipse along the $x$-axis}), $-8^\mathrm{o}$ with the axis $y$ (i.e. the right side closer to the observer) and $-9^\mathrm{o}$ with the axis $z$ (the left part up). In order to reach such an orientation, we follow \citet[][see also their Fig.~1]{2014ApJ...794..174P} and perform successive $x$-$y$-$z$ rotations of our MHD datacubes around respective axes on angles $i\rs{x} = -49^\mathrm{o}$ (the ring plane is horizontal initially in our simulations), $i\rs{y}=-5^\mathrm{o}$, $i\rs{z}=-3^\mathrm{o}$.

The synthesis of polarization maps is based on a method developed by  \citet{2016MNRAS.459..178B,2017MNRAS.470.1156P}. It includes an important effect of the Faraday rotation of the polarization planes along the line of sight in the SNR interior as well as prescription of the turbulent MF evolution downstream of the shock. 
We would like to note that we account completely for the complex internal structures of density and MF inside SNR while calculating the rotation of the polarization planes. 
In the present paper, we adopt this method for synthesis of the Stokes-parameters maps for young SNR in medium with nonuniform distribution of density and MF \citep[details of calculations of the radio emissivity are given in][]{2007A&A...470..927O,2019A&A...622A..73O}. Electrons are assumed to be injected independently of the shock obliquity (this is not a limitation of our model because MF has nearly the same obliquity everywhere in the emitting "barrel" of SN1987A), accelerated on the forward shock and evolved downstream according to the HD structure of SNR. 
The energy spectrum of the radio-emitting electrons is taken to be power-law. We adopt the radio spectral index $\alpha=0.7$ \citep{2010ApJ...710.1515Z} for our images. 

The evolution of the radio flux from the model MOD-B1 agrees with the observed light curve if relativistic electrons are provided by the forward shock only \citep[Fig.~8 in][]{2019A&A...622A..73O}. 
Therefore, we do not consider the particle acceleration at the reverse shock of SN1987A. In fact, we exclude from the image synthesis the cells  {where $\geq 10\%$ of material is ejecta (the number $10\%$ is not fine tuned; in fact any number up to $\simeq 60\%$ produce almost the same results because the contact discontinuity transition is rather sharp)}. Reasons for a minor contribution or suppression of the radio emission from the reverse shock in SN1987A are considered in details by \citet{2019A&A...622A..73O}. Similar effect is found in SN1993J where the high absorption of the radio emission by the ejecta is observed with the ejecta opacity close to 100\% \citep{2011A&A...526A.142M}.

\begin{figure*}
  \centering 
 \includegraphics[width=0.48\textwidth]{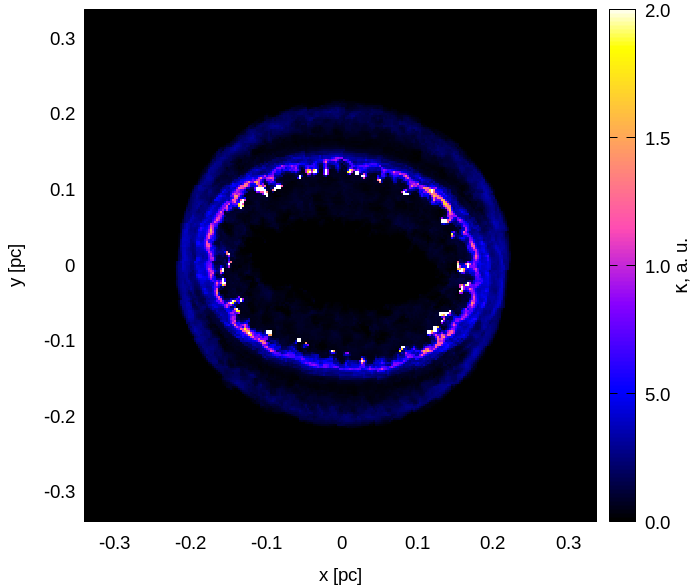}\ \  
  \includegraphics[width=0.48\textwidth]{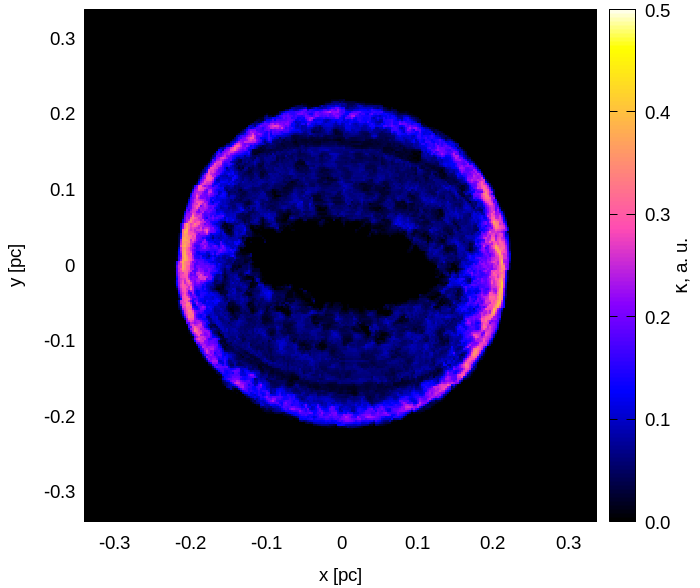}
   \includegraphics[width=0.48\textwidth]{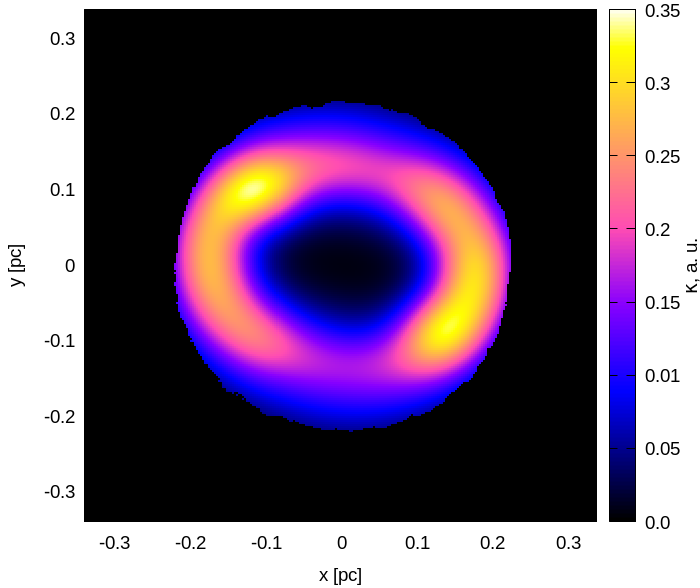}
  \includegraphics[width=0.48\textwidth]{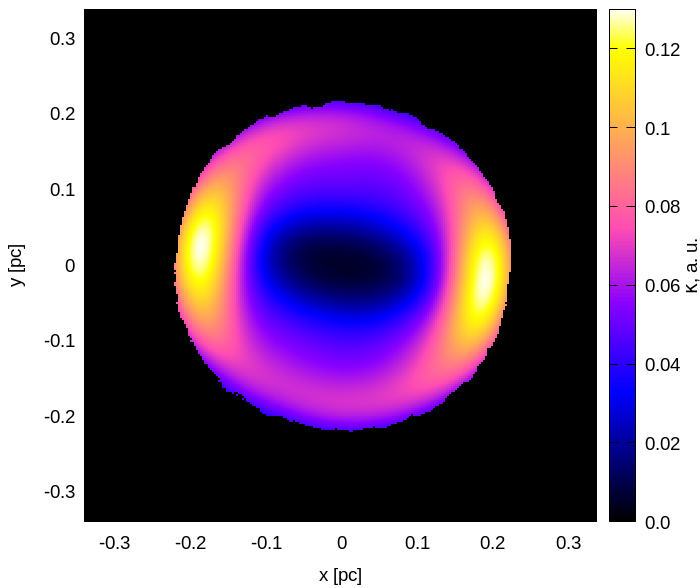}
  \caption{Spatial distribution of the radio-emitting electrons in the two alternatives considered in the present paper. These are the maps of the $\kappa$ integrated along the line of sight (the local synchrotron emissivity is proportional to $\kappa B^{\alpha+1}$).
  {\bf Left}: \textit{alternative A}, including the ring and the clump structures. 
  {\bf Right}: \textit{alternative B}, without these dense components.
  Upper panels: in full resolution, lower panels: smoothed to the resolution of the radio  observations.
  }
  \label{sn87a:sn87kappa}
\end{figure*}

In the present paper, we neglect the turbulent component of MF assuming that the observed polarization pattern is dominated by the ordered component MF. Turbulent MF if randomly oriented everywhere in the SNR interior should not alter the global MF structure which is sampled by the polarization images. Such assumption is in agreement with the observed polarization planes in SN1987A which look quite ordered. It also simplifies considerably production of the polarization direction map that is the main goal toward understanding the global MF structure in SNR. However, this simplification prevents us from producing the polarization fraction map which is sensitive to the MF randomness. We will use an approximate tool in order to estimate the ratio of the random $\delta B$ to the ordered $B$ MF components in SN1987A. 

The number of particles injected and accelerated at the shock expanding into the non-uniform medium -- and therefore the normalization of the momentum spectrum of the radio-emitting electrons $\kappa$ -- is generally considered proportional to the pre-shock number density $n\rs{o}$  \citep{2007A&A...470..927O}. For references, the number densities in the structures surrounded the SN progenitor are: HII region $90\un{cm^{-3}}$, equatorial ring $1000\un{cm^{-3}}$, dense clumps in the ring $2.5\E{4}\un{cm^{-3}}$ on average (see Table~1 in \citet[][]{2019A&A...622A..73O} and Fig.~1 in \citet[][]{2015ApJ...810..168O}). 

Multi-wavelength observations of SN1987A prove the shock passage through the equatorial ring and interactions with clumps during recent decades (e.g. in optical \citep{2008A&A...479..761G,2015ApJ...806L..19F} and X-ray \citep{2013ApJ...764...11H,2012A&A...548L...3M} photons). 
Recent observations even show that the shock has already left the outer edge of the ring 
(\citet{2016ApJ...829...40F,2021ApJ...916...41S,2021ApJ...922..140R} in the X-ray band; \citet{2018ApJ...867...65C} in the radio band; \citet{2019ApJ...886..147L} in the optical band). The period of SNR interaction with the ring is clearly visible in the thermal X-rays:  the light curve and the spectrum are dominated since year 14 after the explosion by the emission from the material of the ring \citep{2015ApJ...810..168O,2019A&A...622A..73O,2020A&A...636A..22O}. In particular, the bump in the X-ray light curve on Fig.~5 in \citet{2019A&A...622A..73O} is proved to be due to the illuminated ring material (the dash-dot line on the figure). 

In contrast, the evolution of the radio flux does not exhibit a bump in the radio light curve (Fig.~5 in \citet[][]{2010ApJ...710.1515Z} or Figs.~7 in \citet[][]{2019A&A...622A..73O}). Therefore, it could be that the dense ring does not contribute much to the radio light curve. Similar conclusion could follow from analysis of \citet{2014ApJ...794..174P}: the discrepancy is evident between the evolution of the observed radio emission and the synthesized one from the HD model. 
The differences which eventually could be due to the same reason are visible from the X-ray and radio images. The brightest regions in the sequence of the X-ray images in the years 2000-2016 seems to be dominated by the dense clumps \citep[Fig.~5 in][]{2016ApJ...829...40F} and an inversion of the East/West asymmetry into the West/East one is evident during these years (Fig.~6 in the same reference). In contrast, the sequence of the radio images \citep[Fig.~1 in][]{2018ApJ...867...65C} has the brightness maximum always at the East during the whole this period. This difference between the radio and X-ray images is also well demonstrated at Fig.~1 by \citet{2014ApJ...782L...2I} for the years 2011-2012.
Again, it seems that the very dense material of the ring and clumps does not manifest itself in the radio band (maybe due to a rapid decrease of injection or acceleration efficiencies, e.g. in portions of the shock which rapidly decelerate in the high-density environment and quickly become radiative). 

Another result also supports the scenario that no significant radio emission originates from the ring. The analysis in the Fourier space by \citep{2008ApJ...684..481N,2013ApJ...777..131N} shows that the geometrical shape of the radio morphology of SN1987A can be a thick torus. In fact, the toy model of a thick torus adopted by these authors fits well in describing the emission originating from the HII region without the need to invoke emission from the shocked ring.

In view of these points, we consider the two alternatives in the present paper. Namely, the \textit{alternative A} keeps the proportionality $\kappa\propto n\rs{o}$ everywhere. The \textit{alternative B} excludes the material of the ring and clumps in calculation of $\kappa$ if the fraction of the ring material in a cell is between 0.3 and 1 and the number density $>1000\un{cm^{-3}}$. Note that this material is present in all other aspects of the model, e.g. in calculation of the internal Faraday rotation. The distributions of the emitting electron density in both these alternatives are shown on Fig.~\ref{sn87a:sn87kappa}. 

\begin{figure*}
  \centering 
  \includegraphics[width=0.48\textwidth]{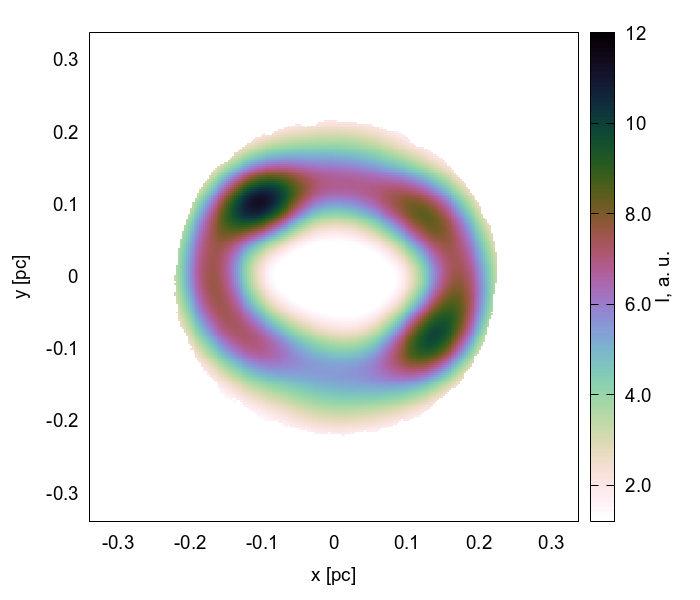}\ 
  \includegraphics[width=0.48\textwidth]{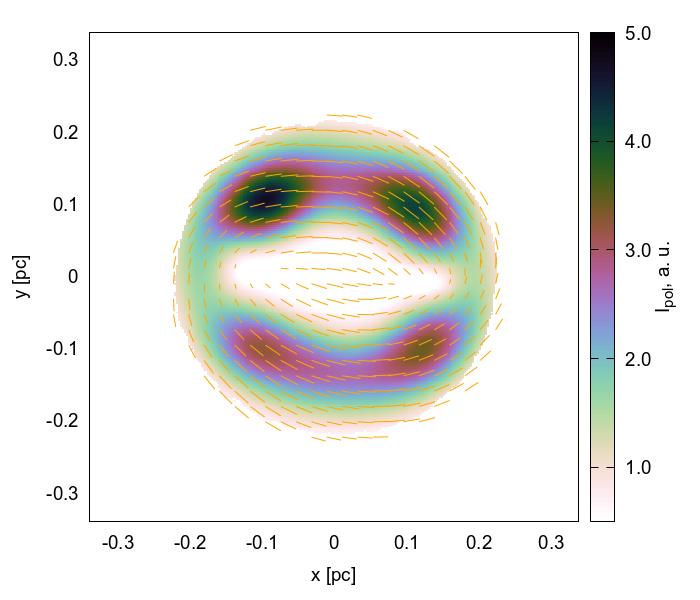}
  \caption{
  Maps of the Stokes parameter $I$ (left) and $I\rs{pol}$ (right)  {for the model MOD-B1 and} for the \textit{alternative A}. 
  Polarization vectors corresponds to MF orientation and are proportional to the polarization fraction.
  Hereafter, our images corresponds to year 30 of SNR evolution, to the frequency 22 GHz and are convolved with Gaussian with FWHM=0.\arcsec 4. 
  Arbitrary units in color scales are the same on Fig.~\ref{sn87a:sn87pol_ring} and Fig.~\ref{sn87a:sn87pol_noring} in order to make it possible to compare the images one to another. 
  }
  \label{sn87a:sn87pol_ring}
\end{figure*}
\begin{figure*}
  \centering 
  \includegraphics[width=0.48\textwidth]{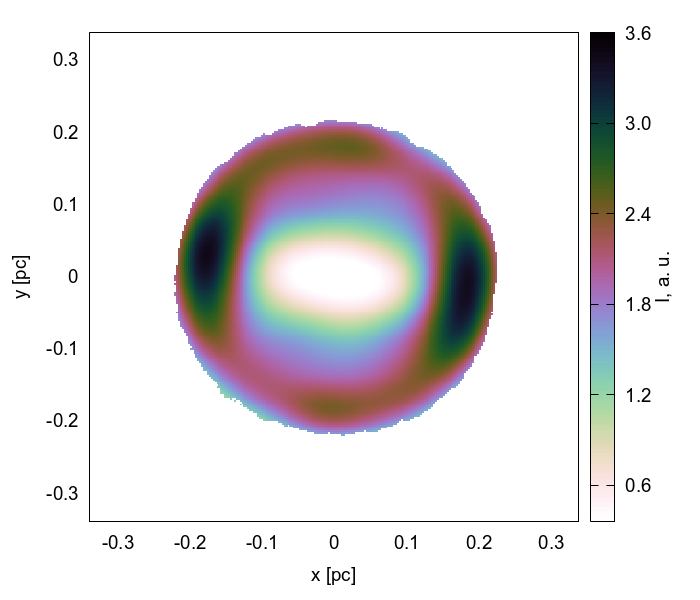}\ \
  \includegraphics[width=0.48\textwidth]{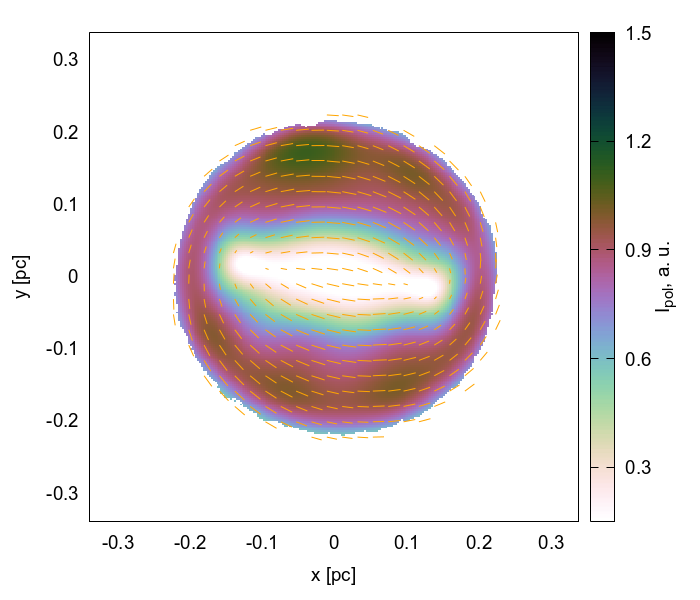}
  \caption{
  The same as on Fig.~\ref{sn87a:sn87pol_ring} for the  for the \textit{alternative B} (without contribution of the ring and clumps to the number density of relativistic electrons).
  }
  \label{sn87a:sn87pol_noring}
\end{figure*}

\subsection{Polarization images for the existing MHD model}

The synthesized images of the total $I$ and the polarized intensity $I\rs{pol}$ 
 {for the model MOD-B1 from \citet{2019A&A...622A..73O}} are shown on 
Fig.~\ref{sn87a:sn87pol_ring} for the \textit{alternative A} and on Fig.~\ref{sn87a:sn87pol_noring} for the \textit{alternative B}. 
The map of the Stokes parameter $I$ on our Fig.~\ref{sn87a:sn87pol_noring} (left) corresponds to Fig.~9 by \citet{2019A&A...622A..73O} but the age, frequency and convolution differ. 
In the present paper, the modelled images of polarized synchrotron emission from SN1987A are reported for the first time.

The following statements are evident.
\begin{itemize}
\item Brightness in the \textit{alternative A} is affected by the location of the clumps. They should effectively influence the radio light curve if they contribute to relativistic electrons. 
\item Radio emission arises in physically different regions of SNR in the two alternatives. 
The dense ring is dominant in the \textit{alternative A} and it should affect the light curve as well. The image in the \textit{alternative B} is mostly from the shocked HII material and the emitting material is shaped like a barrel. 
\item These differences are evident from the synthesised images in full resolution. However, there is no clear way to distinguish between the two alternatives from observations because of  insufficient instrumental resolution. 
\item The shape of the image in the \textit{alternative A} is more elliptical compared to the \textit{alternative B}.
\item The tangential pattern dominates in the distributions of the polarization vectors on Figs.~\ref{sn87a:sn87pol_ring} and \ref{sn87a:sn87pol_noring}. Observed polarization in SN1987A \citep{2018ApJ...861L...9Z} is more radial than it is in the MOD-B1 model. This conclusion is valid for the both alternatives for $\kappa$. 
\item  {Therefore, it is evident that the MOD-B1 model does not reproduce the observed polarization map \citep[reported on Fig.~3 in][]{2018ApJ...861L...9Z}. Thus, MF in SN1987A has structure different from that adopted in this model. }
\item \citet{2019A&A...622A..73O} has also reported numerical MHD simulations for another model, MOD-B100.  {It was shown that this model does not reproduce the radio light curve. In addition,} it has higher tangential MF component  {($A_2=8\E{13}\un{G\ cm}$)}  {comparing to MOD-B1 and therefore} does not correspond to the observed polarization. 
\end{itemize}

 {It has to be pointed out that the configuration of the magnetic field is important for polarization maps, but not for changing the overall dynamics of the SNR which would be analogous even with a configuration of MF which is different from that adopted in our previous models MOD-B1 and MOD-B100.}

In order to check how lower or higher MF strengths or electron density affect the polarisation patterns due to the Faraday effect, we multiplied artificially the rotation measure $RM$ by the same factor everywhere in the SNR interior. The images for different factors in the range 0.3-3 are indistinguishable. 
Though the internal Faraday rotation could change the polarization pattern, it is ineffective for the high frequency of available observations, $\nu=22\un{GHz}$, because $RM\propto \nu^{-2}$. 
The polarization observations at lower frequencies which are sensitive to the Faraday effect would be of a considerable importance for testing the models of MF in SN1987A.

\section{Method for magnetic field reconstruction in HD simulations}
\label{sn87aMF:sect2}

Hydrodynamic model of SN1987A by \citet{2019A&A...622A..73O} agree with wealth of observational data. However, as we have just seen, the structure of MF in this model has higher tangential component comparing to the polarization pattern observed in SN1987A. Therefore, there is a need to look for a MF structure which agree with the radio polarization. The full-scale three-dimensional MHD simulations are not suitable for exploring the parameter space. Therefore, we have developed a semi-analytical approximate method which helps us 
to `reconstruct' MF structure of evolved SNR once its HD structure is known and a model of initial MF around the SN progenitor before the explosion is assumed. Considering different models of the initial MF we may find the polarization configuration which resembles the observed pattern.

\subsection{Description of the method}

Generally, MF pressure in SNRs is considerably smaller than the thermal or ram pressure. SN1987A is not an exception. In the model MOD-B1, for example, the ratio of the thermal to magnetic pressure $\beta>10^{5}$ at its forward shock at the time of its breakout of the stellar surface. Hydrodynamic properties of SN1987A is such that MF at the stellar surface has to be $>1\un{MG}$ in order to affect the overall dynamics of this object. This is far above the values allowed by the radio light curves \citep{2019A&A...622A..73O}.

Considering the limit of the large plasma $\beta$ (i.e. dynamically unimportant magnetic field) one can use HD numerical simulations in order to prescribe the distribution of MF it should be inside a shell-like SNR.

Let the (non-uniform) distribution of MF in the ambient medium before the supernova event $\mathbf{B}\rs{o}(\mathbf{r})$ be known (actually, this is the MF model to be tested). 
In order to recover MF in an element of plasma at time $t$ after explosion, we consider its evolution in this element, in Lagrangian approach. 

At the beginning, we need to know what was MF in this element at the time $t\rs{i}$ when it was shocked. 
The vector $\mathbf{B}\rs{o}$ is split to the two components, the parallel and perpendicular ones 
 ${B}\rs{\| o}=B\rs{o}\cos\Theta\rs{o},$
 ${B}\rs{\perp o}=B\rs{o}\sin\Theta\rs{o}$
where $\Theta\rs{o}$ is the obliquity angle, i.e. the angle between $\mathbf{B}\rs{o}$ and the shock normal $\mathbf{n}$, index `o' marks the pre-shock values. 
If the shock compression ratio is $\sigma$ then the post-shock components are  respectively
 ${B}\rs{\| s}={B}\rs{\|o}$, ${B}\rs{\perp s}=\sigma {B}\rs{\perp o}$;
they lie in the plane fixed by the vectors $\mathbf{B}\rs{o}$ and $\mathbf{n}$.

\begin{figure*}
  \centering 
  \includegraphics[width=0.48\textwidth]{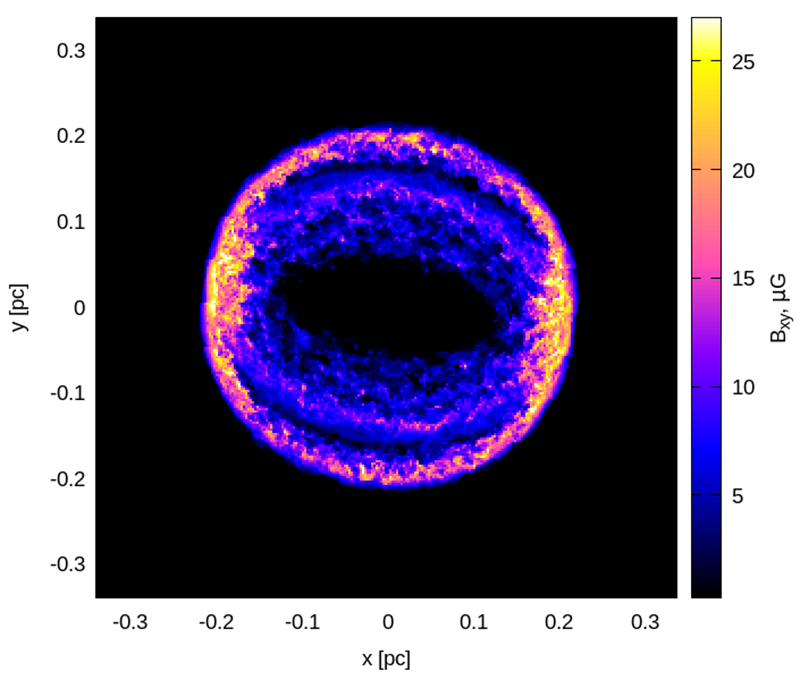}\ \ 
  \includegraphics[width=0.48\textwidth]{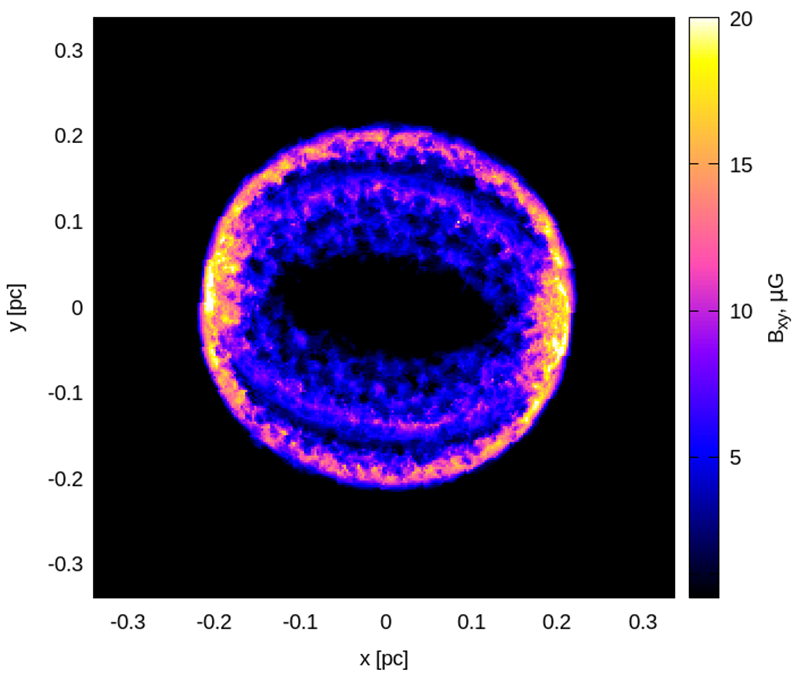}
  \caption{Sum along the line of sight of the MF component perpendicular to the line of sight. 
  {\bf Left}: MF from 3-D MHD simulations.
  {\bf Right}: MF reconstructed from 3-D HD data.
  Note, that these are the MF maps and no distribution of emitting electrons is needed to produce them: MF structure is the same as for \textit{alternative A} and for \textit{alternative B}. 
   {Though the absolute values on the color scales are a bit different, the contrasts between the maximum ($27\un{\mu G}$ on the left, $20\un{\mu G}$ on the right) and the minimum values are the same on both color-scales. In other words, the color-scales differs by a factor only. We have chosen to shift one color-scale on a factor versus the another one in order to emphasise similarities and to make it more suitable to see correlations between the two images. The ratios of brightness between pixels are more important for comparison of images than the difference on the same factor in the overall amplitude.}
  }
  \label{sn87a:sn87mfperp}
\end{figure*}

At any time moment, the obliquity angle in each `shocked' cell may be calculated
assuming that the portion of the shock located at $\mathbf{r}$ at a time moment $t\rs{i}$ runs approximately in the radial direction:
\begin{equation}
 \cos\Theta\rs{o}(\mathbf{a},t\rs{i})=\frac{\big(\mathbf{B}\rs{o}(\mathbf{a})\cdot \mathbf{a}\big)}{|{B}\rs{o}||a|}
\label{sn87aMF:obliq2}
\end{equation}
where $\mathbf{a}\equiv \mathbf{R}(t\rs{i})$ is the Lagrangian coordinate of a given gas element, $\mathbf{R}$ the radius-vector of the shock element. Note, that the three components of the Lagrangian coordinate $\mathbf{a}$ should be kept during the simulations in respective tracers; this allows one to convert $\mathbf{r}$ to $\mathbf{a}$ and vice versa for any time moment $t$. 

A more sophisticated approach could be developed, namely, that the normal $\mathbf{n}$ may not be taken in the radial direction for a moment of time $t\rs{i}$ but parallel to the shock velocity $\mathbf{V}$. 
It could be somehow more accurate than the approximation of the instant radial expansion because it closely follows the paths of the shock elements (respective numerical simulations should  include three additional passive tracers to store the components of the shock velocity vectors). It should be noted however that the velocity of the shock in a non-uniform medium is not always perpendicular to the shock surface. Therefore, such approach would also be approximate. 
The approximation of the radial expansion is simpler and quite accurate for the regions of ambient medium without strong small-scale density gradients which could change considerably the motion direction of a given portion of the shock. 

Magnetic flux through co-moving surface $ds$ is constant in an ideal MHD, i.e. 
$B\rs{\|}ds\rs{\|}=\mathrm{const}$, $B\rs{\perp}ds\rs{\perp}=\mathrm{const}$
. Therefore, 
\begin{equation}
 B\rs{\| s}(a,t\rs{i})a^2=B\rs{\|}(a,t)r^2,\quad 
 B\rs{\perp s}(a,t\rs{i})ada=B\rs{\perp}(a,t)rdr.
\label{sn87aMF2}
\end{equation}  
With the use of the continuity equation $\rho\rs{s}(a,t\rs{i})a^2da=\rho(a,t)r^2dr$, we derive the  expressions:
\begin{equation}
 B\rs{\|}(a,t)=B\rs{\| o}(a)\left(\frac{a}{r}\right)^2,\quad 
 B\rs{\perp}(a,t)=B\rs{\perp o}(a)\frac{\rho(a,t)}{\rho\rs{o}(a)}\frac{r}{a}.
\label{sn87aMF3}
\end{equation}
The formulae for the conversion of $B\rs{\|}(a,t)$ and $B\rs{\perp}(a,t)$ into the three components of $\mathbf{B}(r,t)$ are given by \citet[][Appendix A]{2017MNRAS.470.1156P}.

Thus, the 3-D structure of $\mathbf{B}(\mathbf{r})$ inside SNR depends on the initial model of the ambient field $\mathbf{B}\rs{o}(\mathbf{r})$, on the initial ambient density distribution $\rho\rs{o}(\mathbf{r})$ and on the actual density structure $\rho(\mathbf{r})$ inside SNR. Therefore, introducing ideas for possible distributions of the pre-SN MF $\mathbf{B}\rs{o}(\mathbf{r})$, one can forecast how would MF $\mathbf{B}(\mathbf{r})$ be distributed on-top the 3-D HD structure of evolved SNR. 

Such quasi-MHD model of SNR (3-D data-cubes for HD parameters from the full-scale numerical simulations plus semi-analytically reconstructed data-cubes for MF vectors) may be used to simulate polarization images. These images may then be compared to respective observations. This receipt allows for a rather quick tests of different initial MF configurations $\mathbf{B}\rs{o}(\mathbf{r})$ in cases when the HD structure of SNR known. Once a model for $\mathbf{B}\rs{o}(\mathbf{r})$ which resembles observations is known from this method, then the full-scale MHD simulations may be performed with this initial ambient MF in order to study details of MHD evolution of SNR.

In the present paper, we use this approximate method to realize hints about the MF and the radio emission structures in SN1987A.

\subsection{Test of the method}

First, we have tested our method on the Sedov SNR by comparing reconstructed MF with MHD simulations for the \citet{1959sdmm.book.....S} problem. The reconstructed MF is practically the same as in the 3-D MHD simulations of the Sedov SNR. 
The reason for such accuracy is that our method corresponds to the exact analytical solution for MF profiles downstream of the strong one-dimensional shock from a point explosion in a perfectly conducting gas with weak MF \citep{1964JAMTP...4..113K}. 

Then, we have performed the test of the method with the numerical data for MOD-B1 model by  \citet{2019A&A...622A..73O}. 
Fig.~\ref{sn87a:sn87mfperp} compares MF which contribute to the synchrotron emission, namely, the component perpendicular to the line of sight of MF overtaken by the forward shock. Data from the MOD-B1 numerical model of SN1987A is shown on the left and for the MF reconstructed with our method (with the use of HD data from the same numerical model and with the same $\mathbf{B}\rs{o}(\mathbf{r})$) is shown on the right. We see good correspondence of both MF patterns. 

In summary, the proposed method is adequate even for rather complicate cases like SN1987A if the general MF morphology is of interest. 

\begin{figure*}
  \centering 
  \includegraphics[width=0.49\textwidth]{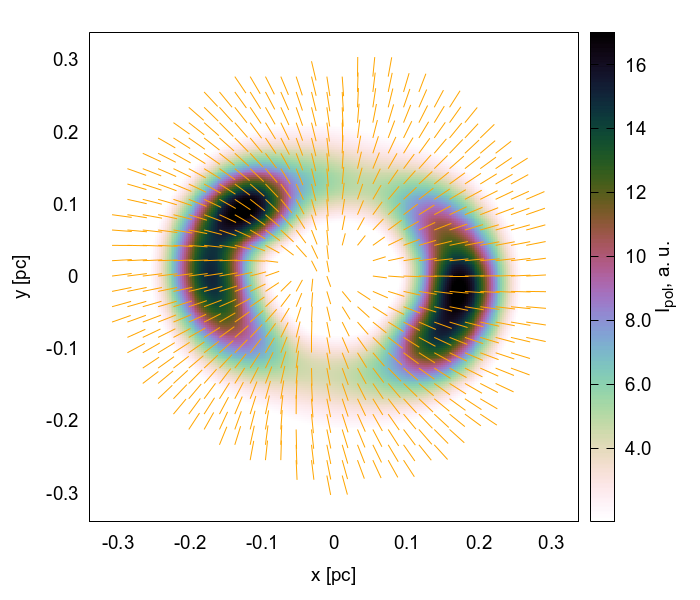}\ \ 
  \includegraphics[width=0.49\textwidth]{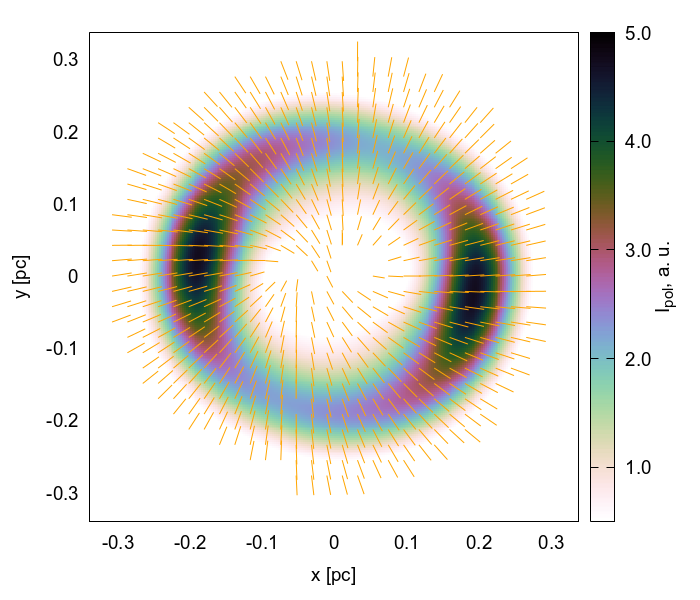}
  \caption{
  Polarization images of SN1987A synthesized by our method for HD data from \citet{2019A&A...622A..73O}. 
  $I\rs{pol}$ is shown by color. Vectors corresponds to the orientation of MF and are proportional to the polarization fraction. 
  {\bf Left}: \textit{alternative A}. {\bf Right}: \textit{alternative B}.
  Arbitrary units in color scales are the same on Figs.~\ref{sn87a:sn87pol_rec_SO2019_1} and \ref{sn87a:sn87pol_rec_SO2019_2} 
  and are $1/100$ of the units for Fig.~\ref{sn87a:sn87pol_ring} and \ref{sn87a:sn87pol_noring}. 
  }
  \label{sn87a:sn87pol_rec_SO2019_1}
\end{figure*}
\begin{figure*}
  \centering 
  \includegraphics[width=0.49\textwidth]{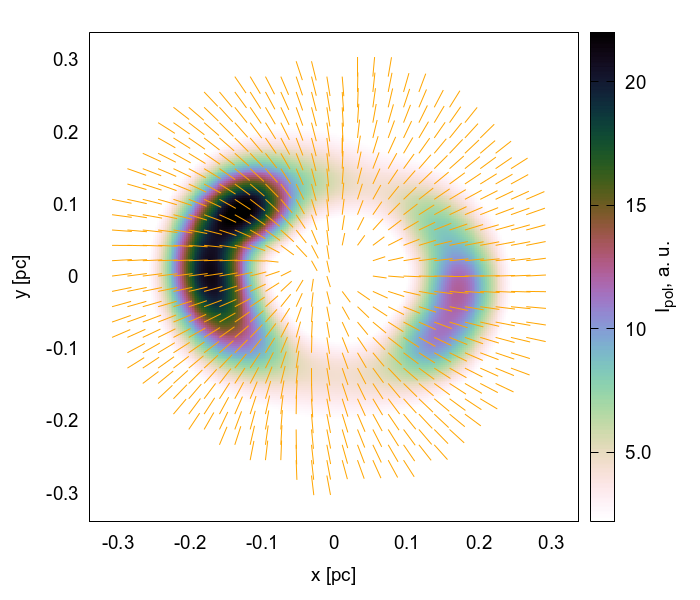}\ \ 
  \includegraphics[width=0.49\textwidth]{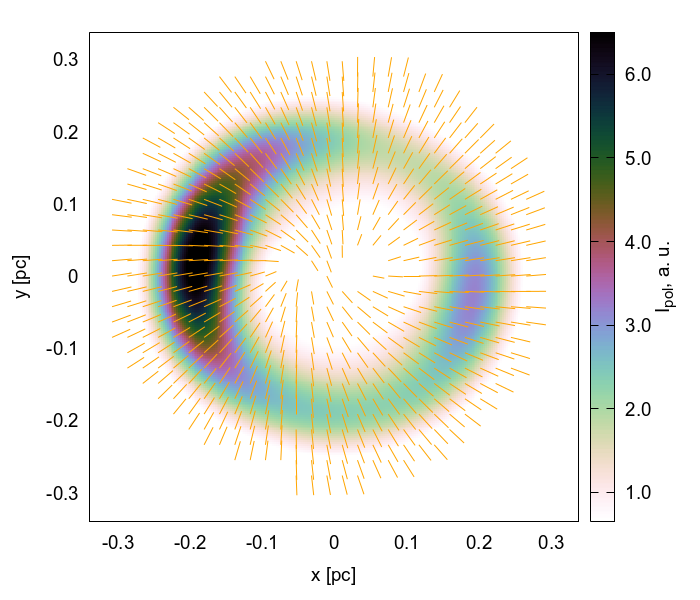}
  \caption{The same as Fig.~\ref{sn87a:sn87pol_rec_SO2019_1} with gradient of emissivity added.
  }
  \label{sn87a:sn87pol_rec_SO2019_2}
\end{figure*}

\section{Results and Discussion}
\label{sn87aMF:sect3}

\subsection{Morphology of the synchrotron emission}

We have used the approximate approach described in Sect.~\ref{sn87aMF:sect2} in order to find an initial MF configuration which results in the polarization image of SN1987A similar to the observational one reported by \citet[][Fig.~3]{2018ApJ...861L...9Z}. 

We applied the method to HD data from the numerical model presented by \citet{2019A&A...622A..73O}. Fig.~\ref{sn87a:sn87pol_rec_SO2019_1} shows the resulting images for  the initial MF given by the equations (\ref{sn87a:parker1})-(\ref{sn87a:parker2}) with the same $A_1$ as in the model MOD-B1 but with a considerably smaller $A_2=2\E{9}\un{G\ cm}$. Such MF results in a required pattern of polarization vectors. Therefore, the stellar rotation  $\omega\rs{s}$ was slower and/or the wind speed $u\rs{w}$ was faster before the SN1987A explosion compared to what was assumed in the MOD-B1 model.

The pattern of the polarization vectors is recovered in our images, in both alternatives for the distribution of the radio-emitting electrons. However, the brightness distribution (both total and polarized) is symmetric on our case, contrary to the observations where the left limb is brighter than the right one \citep[][Fig.~3]{2018ApJ...861L...9Z}. This is not surprising because i) the configuration of the circumstellar medium in the considered numerical model is symmetric (except of the clump locations which are random); ii) the SN explosion was assumed to be spherically symmetric; and iii) the \citet{1958ApJ...128..664P} formulae describe the symmetric circumstellar MF configuration. 

At this point, we do not discuss the nature of the asymmetry but instead perform a simple procedure. Namely, In order to recover the east-west asymmetry in SN1987A image, we have added artificially a gradient of the emissivity by multiplying the `local' values of the Stokes parameters $I$, $Q$, $U$ in each cell of the rotated 3-D cube by a factor $\exp(-x/H)$ with $x$ the Cartesian coordinate and $H=0.5\un{pc}$ the length-scale (to be compared to the radius of SN1987A $0.23\un{pc}$).

The images we derive are shown on Fig.~\ref{sn87a:sn87pol_rec_SO2019_2}. 
They resemble the observed patterns, even in small-scale details. Namely, i) the radial directions of the polarization vectors are dominant; ii) the directions of the vectors over the different parts of the SNR edge are almost the same as in the observational data, even deviations from the radial directions; iii) the sizes of the polarization vectors over the image approximately matches the observed distribution; iv) there are two blind spots around the center of the image; v) the SW-NE direction of the vectors between these two spots is also recovered (it is more clearly visible on Fig.~\ref{sn87a:sn87pol_rec_SO2019_2} right).

\subsection{East-West asymmetry} 

The observed east-west asymmetry in the radio images of SN1987A \citep{2018ApJ...867...65C,2018ApJ...861L...9Z} may have different nature. Since the synchrotron emission is proportional to the product $\kappa B^{\alpha+1}$, the asymmetry should arise mainly from the gradient of magnetic field or density \citep{1997ApJ...479..845G}. Radio images of SNRs are more sensitive to the gradient of MF \citep{2007A&A...470..927O}. In fact, to be solely responsible for the exponential factor in emissivity, the length scale of the non-uniformity for the ambient density has to be  $H=0.5\un{pc}$ while the length scale for the magnetic field strength could be larger, $H(\alpha+1)=0.85\un{pc}$, i.e. the gradient smaller.

X-ray images \citep{2016ApJ...829...40F} also exhibit East-West asymmetry. However, the origin of the X-ray asymmetry could differ from the radio asymmetry because of the thermal nature of X-ray emission which is dominated in the last decade by the dense structures of ring and clumps \citep{2015ApJ...810..168O}. Really, the radio emission is proportional to the density while the thermal X-rays to the density squared. What is also important that the asymmetry in X-rays is inverted: eastern part of SNR was brighter till the year 2007 but then the western part is dominant \citep[Fig.~6 in][]{2016ApJ...829...40F}. In the radio band, the eastern part continue to be brighter \citep{2018ApJ...867...65C}.  {The brightness inversion in the thermal X-rays, together with the fact that the east side faded first in the optical, may be a sign that the west side of the remnant is denser. If so, the magnetic field gradient, directed to the East, should be stronger (i.e. with smaller $H$ than estimated above) to compensate for the density gradient directed to the West.}

Could the dense clumps be (partially?) responsible for such behaviour of the X-ray images? After all, the shock encounters each of them at different times \citep{2015ApJ...806L..19F}. 
Could the asymmetry in the radio images be also influenced by the clumps? 
What is dominant in the asymmetry of the radio images: a large-scale gradient of density or of magnetic field? Which MF component, ordered or disordered, is more asymmetric?  {Could the asymmetry be due to anisotropy of electron injection or acceleration?} 
 {Might it be due to a non-spherical supernova explosion? 
Expanding with different velocities the parts of the shock may reach dense HII region and equatorial ring earlier at one side than at the other. } 
In order to answer these questions, dedicated studies have to be performed. Analysis may benefit from observational data in different electromagnetic domains which may help to disentangle contributions to the emission from density and magnetic field.

\subsection{Radial versus tangential components of magnetic field}
\label{sn87a:rotation}

\subsubsection{ {Consequences from the model of the Parker spiral}}

 {It was demonstrated \citep{2021MNRAS.505..755P} by 1-D MHD simulations that it is unlikely that the radial MF may dominate the tangential one downstream of the forward shock in young SNRs if their strengths were of the same order in the pre-shock medium. The radial MF component drops faster downstream than the tangential component. 

In agreement with this, we have found in the present paper} that the preferentially radial alignment of the polarization vectors in SN1987A should be due to the dominant radial initial MF around the progenitor. 

 {In our study, we adopted the \citet{1958ApJ...128..664P}  model of MF.} 
In the equatorial plane, the ratio between the absolute values of the radial and tangential components  {in this model} varies with a distance and depends on the ratio between the stellar wind speed and the rotation velocity 
\begin{equation}
 \frac{B\rs{r}}{B\rs{\phi}}=\frac{u\rs{w}}{\omega\rs{s}}\frac{1}{r}.
\end{equation}
The period of magnetic rotation is generally considered the same as for the stellar rotation,  $P=2\pi/\omega\rs{s}$. The two periods may differ if MF is not frozen into the outer layers of a star. 

In the model MOD-B1 \citep{2019A&A...622A..73O}, $u\rs{w}=500\un{km/s}$ and $P=2\E{4}P_\odot$ with $P_\odot=27\un{days}$ were taken. For these values, the ratio $B\rs{r}/B\rs{\phi}=0.52$ at a distance $r=0.23\un{pc}$ (the radius of SN1987A). Thus, the tangential component is high in the MOD-B1 model. 

Our results shows that the observed polarization pattern may be reproduced if the radial MF component  {in the \citet{1958ApJ...128..664P} MF model} is dominant on the length-scale of the radius of SN1987A, $B\rs{r} \gg B\rs{\phi}$. Numerically, the ratio should be $B\rs{r}/B\rs{\phi} \gtrsim 10$ at $r=0.23\un{pc}$. This may be reached if 
$u\rs{w}/\omega\rs{s}$ in expression (\ref{sn87a:parker2}) is more than $\sim 20$ times larger (and thus $A_2$ smaller) than in the MOD-B1 model. 
This factor 20 is a sort of a lower limit (corresponding upper limit to $A_2$ is $\sim 4\E{9}\un{G\ cm}$): smaller value of the factor or larger SNR radius will cause more tangential polarization pattern.\footnote{The ratio ${u\rs{w}}/{\omega\rs{s}}$ yields the radius in the equatorial plane $r_0$ where the tangential velocity due to the rotation $\omega\rs{s} r_0$ equals to the radial velocity of the wind $u\rs{w}$. At this radius, $B\rs{r}=B\rs{\phi}$. The tangential MF component gets progressively higher than the radial one for distances $>r_0$ and, at later times (when SNR radius becomes larger) the SNR feels mostly the $B\rs{\phi}$ component. For reference, $r_0=0.12\un{pc}$ in the MOD-B1 model  {and should be larger than $2.4\un{pc}$ to be consistent with the radial polarization in our model of SN1987A}.}

Our images reported above were synthesized for the factor 40. Though the factor 20 results in the synthesized image which in general is similar to the observed one, the value 40 provides more accurate correspondence to observations on the smaller scales.
If we divide the factor $40$ evenly between the wind and rotation, we come to the numbers $u\rs{w}=3200\un{km/s}$ and $P=9400\un{yrs}$. 
Such wind velocity for the progenitor of SN1987A (B3 class star with temperature $16\un{kK}$) is acceptable. For stars in the main sequence, it varies from $\sim 10\un{km/s}$ for cool to $\sim 3000\un{km/s}$ for hot luminous stars \citep{2015A&A...577A..27J}. It could be in the range $2500-3500\un{km/s}$ for massive stars with temperatures $25-40\un{kK}$ \citep[][their Fig.~2]{2018A&A...619A..54V}. 
In contrast, the rotational period $\sim 10^5P_\odot$ is out of the values reported in the literature. Typically measured periods range from hours to months for stars of different types. There are Ap stars with determined periods of $10-30\un{yrs}$ \citep[e.g.][]{2019A&A...624A..32M,2022A&A...660A..70M} and $\sim 100\un{yrs}$ with expectation that periods may reach several centuries \citep{2017A&A...601A..14M}. 
As stated by \citet{2019A&A...624A..32M}, 
our knowledge about long periods of star rotation is quite incomplete. This is in particular due to difficulties in determination of variations on such long time-scales and in measurement of the radial velocity component $v\sin i$ below the micro-turbulent velocity $\sim 20\un{km/s}$ \citep{2013A&A...559L..10S,2014A&A...562A..37M}. For the sake of comparison, the values needed to be derived in observations are 
$v\sin i=0.004 P_{*}^{-1}\un{km/s}$ for periods $P_{*}$ in units of 1000 yrs and stars with the radius and inclination as in the progenitor of SN1987A.

\subsubsection{ {On the rotation of the SN1987A progenitor}}

 {The spectra of Sanduleak $-69\  202$ observed a decade before the explosion \citep{1989A&A...219..229W} suggest that the rotational velocity of this star may not be above $100\un{km/s}$ \citep{2006BASI...34..385P}. So, the rotational period exceeded $2\un{weeks}$ prior the explosion. It is unknown however to what degree it was larger than few weeks.}

 {The slow rotation of the progenitor star of SN 1987A suggested by this study seems to be in tension with current models of the progenitor. One of the reasons why the single-star progenitor model for SN1987A was rejected is a low rotation rate of a single star  \citep{1992PASP..104..717P}. The binary merger model was introduced because the rotation helps an evolutionary model for a SN1987A progenitor to pass a series of necessary tests \citep{1992PASP..104..717P,2007Sci...315.1103M}. 

The ambient medium where SN1987A expands is axisymmetric and strongly asymmetric in respect to the polar angle. The HD model \citep{2006MNRAS.365....2M,2007Sci...315.1103M,2009MNRAS.399..515M} adopts the binary merger, the Red Supergiant (RSG), the Blue Supergiant (BSG) sequence with strong rotation in order to explain such structure. The MHD model \citep{1996PASJ...48...23W,2002Sci...296..321T} explains the asymmetry by an evolutionary sequence of the RSGF and BSG winds controlled by a strong toroidal magnetic field and also needs a strong rotation as an origin of such MF. 
}

 {Instead, the 
model of the Parker spiral for the pre-explosion ambient MF configuration
together with the radial polarization pattern in SN1987A imply a slow rotation of the magnetic field around the progenitor star. At this point, we may mention the two possibilities when the strong rotation and the radial MF may co-exist in the evolutionary scenario for the progenitor}.

 {First, one might imagine a separation of the ambient magnetic field from a stellar rotator in {\em space}, by decoupling the spinning core from the outer layers of the star which have low angular momentum (M.A.Aloy 2022, private communication). We cannot say anything more at this point because the core-envelope rotational coupling in stars is an  open problem \citep[e.g.][]{2021PhRvD.103f3032S}. Some authors support effective decoupling in terms of rotation between the stellar interior and the outer layers in the isolated single post-main sequence stars \citep{2000MNRAS.315..543H}. Others, considering binary systems, disfavor models with mild rotational coupling \citep{2020A&A...636A.104B}.}
 {Alternatively, deviations from ideal MHD conditions may split the MF and star rotations. MF may not be frozen in the stellar material if there is a non-zero resistivity. In such circumstances, their periods of rotation may differ.}

 {In the second scenario, the radial ambient magnetic field may be separated from the rotating star in {\em time}. The ambient magnetic field at some distance from the star reflects the rotation of that star at the time when this `portion' of the field was produced.
So, when did the star have rotation velocity that low to produce MF configuration with a negligible toroidal MF component? The polarization map we observe now is the result of the shock interaction with the HII region which is believed to be a material of the wind during the RSG stage of the star evolution. Therefore, the slow rotation might be a feature of this phase.
In fact, it is known from the angular-momentum conservation that when the radius of a main-sequence star increases to a supergiant one, the angular rotation of the star decreases considerably \citep{1989ApJ...341..867C}.

The slow-merger scenario suggests that the spinning-up time for the merger, common envelope and mass ejection from the merger is short ($\sim 100\un{yr}$) comparing to the next RSG stage (few $1000\un{yr}$) \citep{2007Sci...315.1103M,2018MNRAS.473L.101U}.
The matter from the merger was ejected in the polar directions \citep{2007Sci...315.1103M,2009MNRAS.399..515M} and takes a large fraction of the angular momentum away \citep{2018MNRAS.473L.101U}. 
Therefore, the `layer' of the ambient medium filled with the material which has the high angular momentum (and thus the tangential MF component) should be rather thin and form polar lobes. 
As a result, it should be inefficient in modification of the overall polarization pattern of SN1987A which is due to emission from the equatorial HII region. 

In the scenario of \citet{2006MNRAS.365....2M,2007Sci...315.1103M,2009MNRAS.399..515M}, 
one of the stars in the initial binary system was already RSG with `slow' (few $10\un{km/s}$) spherically symmetric wind. 
The wind of RSG after the merger phase was within the equatorial polar angles and served as a channel for losing the remaining angular momentum. At the next BSG stage, the wind was spherically symmetric (i.e. non-rotating) and `fast' (few $100\un{km/s}$) in this model.
In other words, there is a change in the regime of the matter outflow from the post-merger RSG to BSG.\footnote{Instead, the Parker MF model assumes the steady wind and rotation.} Namely, the rotation of RSG slows down to the end of RSG epoch and then the material of RSG wind is blown out by the non-rotating BSG wind.  
The inner boundary of the HII region corresponds to the transition from the dense RSG wind to the thin BSG wind. 
Therefore, the MF close to the inner boundary of the HII region (where the SNR shock is propagating now and where the synchrotron emission arises) was originated from a star which was a slow rotator at that time. Afterwards, the matter with this MF was pushed away by the BSG wind to the present-day location. In this process, the angular momentum conservation reduces the (already low) tangential MF even more.}

 {One might suggest a hypothetical way to put limitations on the rotation rate of the star at the time when the equatorial ring was created. 
This dense equatorial ring is believed to be ejected at the RGB phase \citep{1993ApJ...405..337B,2009MNRAS.399..515M}. The location of the ring within the HII region \citep{1995ApJ...452L..45C} supports this hypothesis. 
There are detailed Hubble Space Telescope observations of the ring with the dense blobs resolved \citep[e.g.][]{2015ApJ...806L..19F,2019ApJ...886..147L}. The small changes in their radial locations are detected \citep[Fig.~3 in][]{2015ApJ...806L..19F}. It could be important to derive limitations on the rotational period at the RSG phase by looking for an azimuth shifts of the bright blobs in the equatorial ring.  However, the detected radial motion of clumps ($\Delta R\sim 0.01\un{arcsec}$ on the time-scale $\sim 1\un{yr}$) are due to  velocities of order of few $100\un{km/s}$. 
The azimuth velocity at the radius of the equatorial ring ($r\rs{rg}=0.18\un{pc}$) would be just $R\rs{rsg}/r\rs{rg}\sim 3\E{-5}$ of the equatorial rotational velocity of the progenitor at the time when the ring was created (we have taken $R\rs{rsg}=240\un{R\rs{\sun}}$ as the RSG radius for this estimate). 
Therefore, the azimuth shifts in the ring may be noticeable on $\sim 10\un{yr}$ time scale only if the initial equatorial velocity was of the order of the light speed that is impossible. We may interpret this in the other way. Since the equatorial velocity of the star was smaller than the critical velocity ($\sim 100\un{km/s}$) at the end of the RSG phase (and we are unable to detect today the azimuth speed in the ring), the tangential component of frozen-in MF may not be detected at the present radius of SN1987A.}

\subsubsection{ {Radial polarization patterns in young SNRs}}

 {The preferentially radial orientation of the polarization vectors seems to be common feature for young SNRs while the tangential vectors happen mostly in evolved SNRs  \citep{1976AuJPh..29..435D,2015A&ARv..23....3D}. 
Radial polarization is observed in young SNRs with progenitors of different nature, for example, SN1987A was peculiar type II supernova, Cas~A the type IIb, SN1006 the type Ia. 
Different causes may act in different SNRs but the common property stimulates thoughts about a universal reason for the property. A number of hypotheses for the radial polarization patterns have been suggested. 

Our results reveal that the small tangential MF component in SN1987A could be a property of the progenitor star at some stage of its evolution. 
} 

\begin{figure}
  \centering 
  \includegraphics[width=\columnwidth]{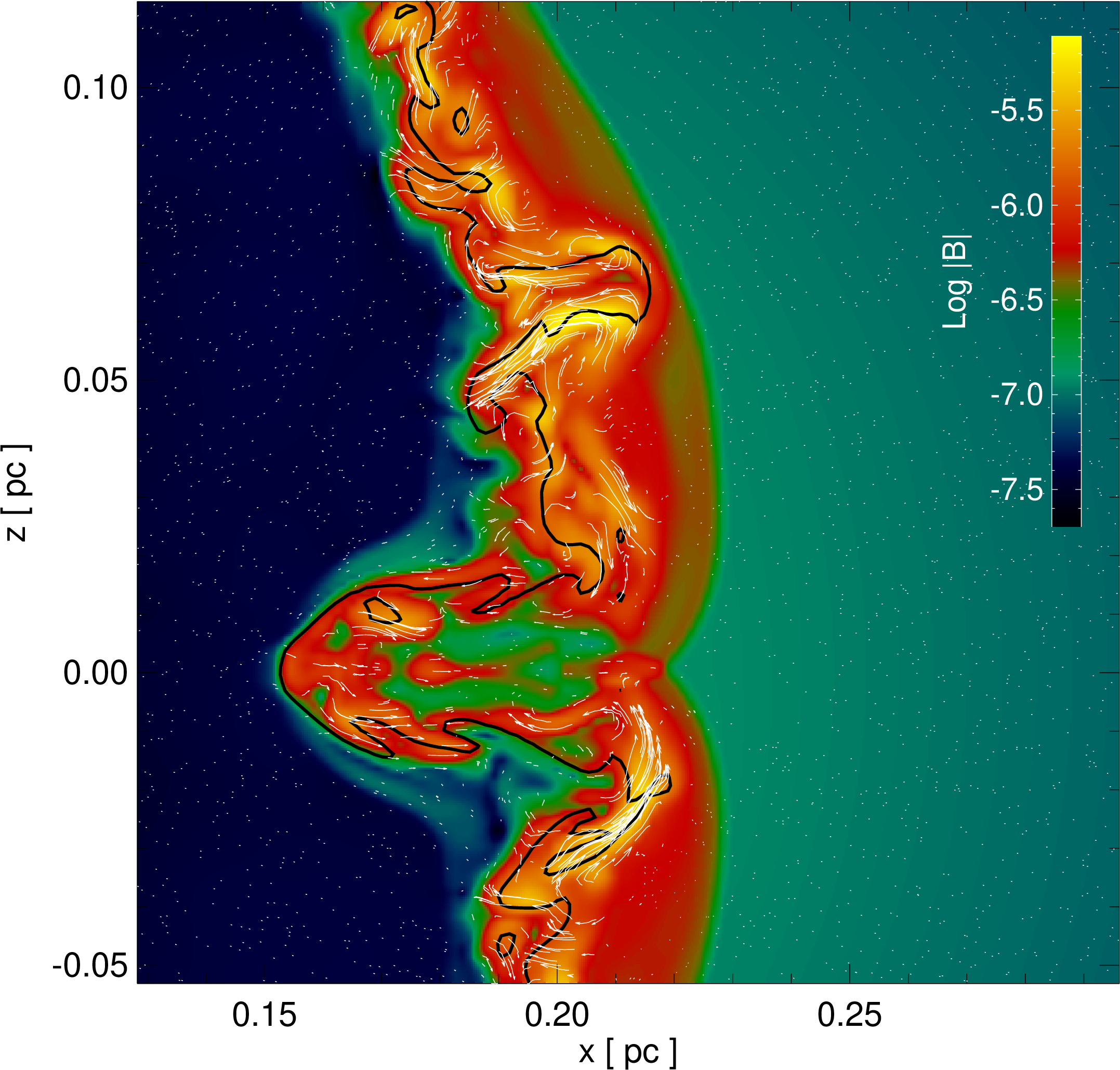}
  \caption{Fragment of a cross section of MF data cube from MOD-B1 model. 
  Colors show the MF strength  {in G}. The MF lines are white. 
  The contact discontinuity is derived from the tracer of ejecta with the threshold at 0.5 and is shown by the black line. The RT finger is located nearly along the coordinate $z\approx 0.06\un{pc}$. The feature along the coordinate $z=0\un{pc}$ is due to the interaction with the equatorial ring.
  }
  \label{sn87a:rtfinger}
\end{figure}

 {The origin of the radial polarization patterns due to the Rayleigh-Taylor (RT) instabilities was considered by  \citet{1996ApJ...465..800J,1996ApJ...472..245J}. The authors concluded in the latter  paper that  RT instability may be responsible for the radially-oriented MF lines in the mixing region around the contact discontinuity.} 
In our 3D simulations \citep{2019A&A...622A..73O}, the RT  instabilities are well developed  {at the border between the swept-up ambient material and the stellar ejecta}. As an example, one quite prominent `finger' is shown on Fig.~\ref{sn87a:rtfinger}. It is elongated along the SNR radius. MF is hydro-dynamically amplified up to 50 times the pre-shock value (yellow color) and has radial direction (MF lines are almost along the radius) there. However, as we see from Fig.~\ref{sn87a:sn87pol_ring} and \ref{sn87a:sn87pol_noring}, the RT instabilities are ineffective to provide the dominant radial orientation of the polarization pattern in  {MOD-B1 MHD model because their `filling factor' is not enough to dominate the polarization pattern}.\footnote{ {It merits to be noted that we adopted the re-grid technique in 3D MHD simulations in order to follow the SNR evolution over orders of magnitudes in space and time \citep{2019A&A...622A..73O}. At each re-grid, we lost information about features on the smallest length scales. The grid size however is re-scaled to just 1.2 times at each re-grid. The small value in $20\%$ has been chosen for re-grid in order to keep the small features evolving in our simulations. Therefore, we expect that most of RT instabilities are preserved in the course of our simulations.}}  {This is in agreement with the earlier findings, namely, with the low amount of MF amplified by such instabilities \citep{1996ApJ...465..800J}.}

Our approach assumes that the polarization pattern arises in SN1987A due to a dominant contribution from the `ordered' large-scale magnetic field (including RT instabilities) and not due to the `disordered' MF turbulence which could be excited by accelerated cosmic rays at the smaller scales  {(we cannot treat it because of numerical limitations imposed by the available  spatial resolution)}. A turbulent MF in SNRs may also be originated by the Richtmyer–Meshkov instability 
induced at the forward shock by its interaction with the pre-existing upstream density perturbations \citep{2013ApJ...772L..20I}. 
 {If shock-accelerated cosmic rays or the shock interaction with the large-scale ambient turbulence} could amplify preferentially the radial MF component and such a process is able to provide the dominant radial field in the most of emitting volume \citep[like considered by][]{2014ApJ...794..174P} then the possibility to infer information about the configuration of the pre-supernova magnetic field from the polarization observations could be questionable.

 {Another possibility for the radial polarization is presented by \citet{2017ApJ...849L..22W}. It is suggested that the pattern could be due to specific distribution of radio emitting electrons. If the particles are accelerated preferentially on the parallel portions of the shock then the only radial component of the completely turbulent MF will be  highlighted. Such effect requires extremely selective particle acceleration and a dominance of the highly disordered MF over the `ordered' component.}

\subsection{Disordered component of magnetic field}

As in the observational results, the vectors on our images are proportional to the polarization fraction. Their lengths are similar to those in the observational image. This means that the polarization fraction maps are also similar. However, the values of that fraction in our model are higher than observed because we excluded for simplicity the turbulent MF component from our consideration.

The mean fraction $\Pi$ of polarized emission from SN1987A at 22 GHz is $2.7\%\pm0.3\%$ in the brightest eastern lobe and $3.6\%\pm1.5\%$ in the inner part \citet{2018ApJ...861L...9Z}. The maximum local $\Pi$ is about $6-8\%$, as from the sizes of vectors in the observational image. 

The polarization fraction decreases in those regions of the SNR projection where MF rapidly changes orientation along the line of sight inside SNR, either due to MHD instabilities or due to turbulent MF. In our 3D model, the instabilities develop and lower the fraction of polarization.
In the MOD-B1 data, the fraction varies over the SNR projection from few per cent to about $50\%$. It is $15\%-25\%$ in the bright eastern limb. 
In order to have an idea about the random MF component in SN1987A, we adopt the approach developed in \citet{2016MNRAS.459..178B,2017MNRAS.470.1156P}. Namely, 
we add to each cell the random MF distributed with the spherical Gaussian with the standard deviation $\delta B$. 
To be more specific, we calculate the local Stokes parameters $Q$ and $U$ with formulae (19)-(20) from \cite{2017MNRAS.470.1156P}, i.e. multiplying  the expressions in the parentheses (which are sensitive to the ratio $\delta B/B$) to the local $Q$ and $U$ which are calculated with the ordered MF which is given by MHD simulations. 
We keep the ratio $\delta B/B$ the same in the whole volume of SNR and synthesize the polarization maps for MOD-B1 model and different values of this ratio. Note, that since we consider the spherical Gaussian for the random MF orientations, the patterns of the Stokes parameters remain the same (i.e. they are determined by the ordered MF component in this approach). However, the values of $\Pi$ decrease with increasing $\delta B/B$. 

\begin{figure}
  \centering 
  \includegraphics[width=\columnwidth]{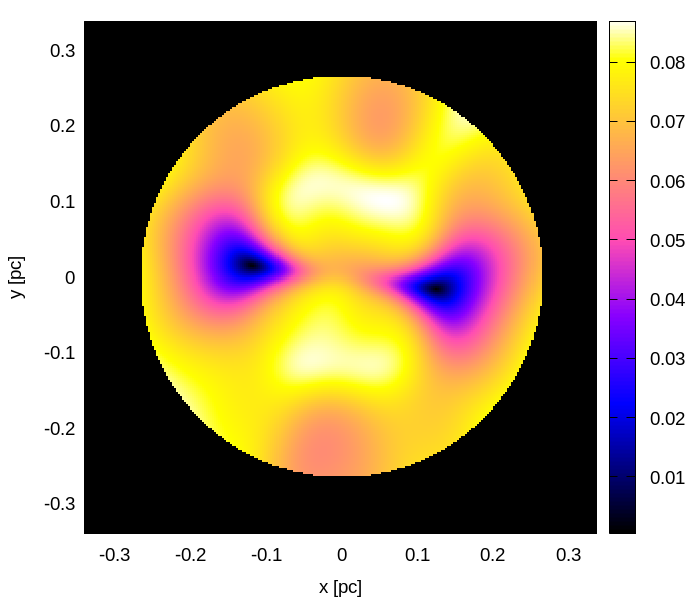}
  \caption{Image for the polarization fraction $\Pi$. MOD-B1 model by \citet{2019A&A...622A..73O} with the random MF added artificially in each cell with ${\delta B}/{B} = 1.5$.
  \textit{Alternative B} of the distribution of emitting electrons. 
  }
  \label{sn87a:sn87dB2}
\end{figure}

The map of polarization fraction with values of $\Pi$ similar to those observed is shown on Fig.~\ref{sn87a:sn87dB2}. It is produced for the $\delta B/B= 1.5$. Thus, we expect that the disordered and ordered MF components in SN1987A are of the same order.

\section{Conclusions}
\label{sn87aMF:concl}

In the present paper, we have modeled the pattern of polarization vectors in SN1987A which resembles the observed one reported by \citet{2018ApJ...861L...9Z}. 
This was done by using an approximate method for reconstruction of the spatial structure of MF in SN1987A at the age of 30 years. 
The method utilizes the numerical 3-D hydrodynamic model of the SN1987A \citep{2015ApJ...810..168O,2019A&A...622A..73O} which agrees with temporal and spectral evolution of the remnant in X-rays. 
We have revealed some properties of magnetic field and structure of the synchrotron emission in SN1987A which may be used for the future full-scale 3-D MHD simulations. 
Namely, an almost radial ambient magnetic field is needed to reproduce the observations. 
If the circumstellar magnetic field in the pre-SN era may be described with a Parker spiral, it should be with a negligible tangential component on the length-scale of $0.23\un{pc}$, the present-day radius of SN1987A. 
From the Parker model  {(which assumes the steady wind and steady rotation)}, the implication would be either quite slow rotation of the progenitor star and/or the very fast stellar wind before the SN explosion. 

 {We pointed out in Sect.~\ref{sn87a:rotation} that the rotation and wind are not steady in the \citet{2007Sci...315.1103M} model, namely, 
i) the star rotation slowed down toward the end of RSG phase, ii) the RSG wind bubble was blown up by the BSG wind. Therefore, the azimuth velocity of the material of the RSG wind has considerably decreased due to the conservation of angular momentum, as also the tangential component of frozen-in MF in SN1987A.}

We consider two scenarios for the electron injection into acceleration process. Namely, alternative B excludes the possibility for electrons to be injected in the very dense material of the equatorial ring and the clumps. In the alternative A, the injection is considered proportional to the local density also there. We see that the synthesized radio image is more elliptical in alternative A (because the contribution from the ring material is high, Fig.~\ref{sn87a:sn87kappa} left) and more circular in the alternative B (because the image is formed by the emitting `barrel', Fig.~\ref{sn87a:sn87kappa} right). The radio emissivity is proportional to $\propto \kappa B^{(s+1)/2}$. Thus, the radio image depends on the relative contributions from the structures of $\kappa$ (electron injection) and of $B$. Therefore, the MF strength is not just a normalization for the radio image of SN1987A but serves also as a `weighting factor' between the alternatives A and B. For example, high MF may be dominant in the formation of the radio image and diminish the role of the dense ring in image for the alternative A. This point has to be taken into account in the future modelling. Being considered together with simulation of the radio light curve, it may put limitations on the MF strength.

An important property of the radio image of SN1987A is the East-West asymmetry in surface brightness \citep{2018ApJ...867...65C}. This feature is also present in the map of the polarized intensity \citep{2018ApJ...861L...9Z}. Various reasons may produce it, in particular, the gradient of ambient density or MF strength, an asymmetry of SN explosion, the anisotropy in cosmic ray acceleration or in MF amplification. 
We have adopted the numerical model of SN1987A with a highly asymmetric explosion  \citep{2020A&A...636A..22O} and demonstrated that it does not result in the observed radio asymmetry.
In order to disentangle other possibilities, dedicated numerical simulations should be performed. It is also important to observe the effect of the internal Faraday rotations. 
Present observational data are taken at frequency where the internal rotation is ineffective in modification of the polarization pattern. 
In order to test alternative hypothesis, polarization observations at longer radio wave-lengths would be desirable. In particular, SKA1 could reach 0.4'' resolution at 1.4 GHz\footnote{SKA Technical information. The Telescopes (retrieved on 2022.06.29)  \url{http://www.skatelescope.org/wp-content/uploads/2018/08/16231-factsheet-telescopes-v71.pdf}}.
In an unclear situation, we added \textit{the gradient of the radio emissivity} into our model. It is shown that observed East-West contrast in the polarized intensity is comparable to the exponential distribution with the length-scale of order of the remnant diameter.

\section*{Acknowledgements}

The work has been partially granted by the Project HPC-EUROPA3 (INFRAIA-2016-1-730897), with the support of the EC Research Innovation Action under the H2020 Programme; in particular, OP gratefully acknowledges the computer resources and technical support provided by CINECA and hospitality and support of Astronomical Observatory in Palermo.
OP and VB thank the Armed Forces of Ukraine for providing security to finalize this work. 
SO, MM, FB acknowledge financial contribution from the
PRIN INAF 2019 grant `From massive stars to supernovae and supernova
remnants: driving mass, energy and cosmic rays in our Galaxy'.
SN and MO are supported by Pioneering Program of RIKEN for Evolution of Matter in the Universe (r-EMU). SN is also supported by JSPS KAKENHI (A) Grant Number JP19H00693.
SL acknowledges financial support from the Italian Ministry of University and Research -- Project Proposal CIR01\_00010. We thank Miguel Ángel Aloy Torás for useful discussions.

\section*{Data availability}

The data underlying this article will be shared on a reasonable request.

\bibliographystyle{mnras}
\bibliography{mfsn1987a}


\bsp	
\label{lastpage}
\end{document}